\documentclass[runningheads]{llncs}
\usepackage{clef25-longeval-invited}

\begin{document}
\title{\fontsize{11.4}{15}\selectfont Simplified Longitudinal Retrieval Experiments:\texorpdfstring{\\}{ } A Case Study on Query Expansion and Document Boosting}
\titlerunning{Simplified Longitudinal Retrieval Experiments}

\author{
    Jüri Keller\inst{1} \orcidID{0000-0002-9392-8646} \and
    Maik Fr{\"{o}}be\inst{2} \orcidID{0000-0002-1003-981X} \and
    Gijs Hendriksen\inst{3} \orcidID{0000-0003-0945-3148} \and
    Daria Alexander\inst{3} \orcidID{0000-0001-9478-7083} \and
    Martin Potthast\inst{4} \orcidID{0000-0003-2451-0665} \and
    Philipp Schaer\inst{1} \orcidID{0000-0002-8817-4632} 
}

\authorrunning{Keller et al.}

\institute{
    TH Köln - University of Applied Sciences, Cologne, Germany \and
    Friedrich-Schiller-Universität Jena \and
    Radboud Universiteit Nijmegen \and
    University of Kassel, hessian.AI, ScaDS.AI
}

\maketitle
\begin{abstract}
The longitudinal evaluation of retrieval systems aims to capture how information needs and documents evolve over time. However, classical Cranfield-style retrieval evaluations only consist of a static set of queries and documents and thereby miss time as an evaluation dimension. Therefore, longitudinal evaluations need to complement retrieval toolkits with custom logic. This custom logic increases the complexity of research software, which might reduce the reproducibility and extensibility of experiments. Based on our submissions to the 2024~edition of LongEval, we propose a custom extension of \irdatasets for longitudinal retrieval experiments. This extension allows for declaratively, instead of imperatively, describing important aspects of longitudinal retrieval experiments, e.g., which queries, documents, and/or relevance feedback are available at which point in time. We reimplement our submissions to LongEval~2024 against our new \irdatasets extension, and find that the declarative access can reduce the complexity of the code.


  \vspace{1em}

  \parbox[c]{\columnwidth}
  {
    \hspace{1.1em}
    \faGithub \ \href{https://github.com/clef-longeval/ir-datasets-longeval}{\nolinkurl{https://github.com/clef-longeval/ir-datasets-longeval}}}
\end{abstract}

\section{Introduction}
Information Retrieval (IR) is centered on the task of finding relevant information to meet users' information needs. Therefore, systems must be able to cope with the ongoing flood of information. Current IR test collections abstract this complex task to enable a well-proven evaluation framework. This abstraction excludes certain aspects, such as temporal dynamics, that are inherent to real-world retrieval environments and tasks.
The LongEval lab reintroduces some of these dynamics into offline evaluations with dynamic test collections~\cite{DBLP:conf/clef/AlkhalifaBBCDEASGGKLLMPPSMZ23,DBLP:conf/clef/AlkhalifaBDEAFGSGILMMMP24,DBLP:conf/ecir/CancellieriEFGSGIKKMPPS25}. That is, a changing document corpus, evolving information needs, and updated relevance judgments are considered. This makes the evaluations more realistic in these regards and allows new directions for retrieval approaches. For example, if previous relevance judgments are available, based on user interactions, as in the LongEval test collections, they can be used as a relevance signal. Our previous works submitted to the lab have shown that such signals can strongly improve retrieval effectiveness at low cost~\cite{alexander:2024,keller:2024,keller:2025a}.

While IR systems are exposed to changing data, users expect consistent, good effectiveness. To provide and maintain this, it is essential to regularly assess the system. Such longitudinal evaluations substantially increase the complexity compared to conventional Cranfield-style offline evaluations. With each new snapshot of a search setting, new versions of documents, queries, and qrels are introduced that need to be stored and maintained. Additionally, the retrieval approach and adequate baselines must take care to process only past data, while it would be helpful if they can easily process new and modified versions of the data. The number of necessary experiments increases with each variation of the parameters and snapshots. That means that the complexity, the demand for resources, and the propensity to errors of the experiments drastically increase.

These challenges of longitudinal evaluations make software submissions, as enabled by TIRA~\cite{DBLP:conf/ecir/FrobeWKGELHSP23}, difficult, as the submitted software becomes more complex and, as a result, the maintenance effort for organizers would increase (e.g., because every submission would come with different custom code to load the data). Still, in longitudinal settings, software submissions are especially interesting because they allow the application of the exact same approach to different snapshots. 

To simplify longitudinal retrieval experiments and thereby facilitate software submissions in longitudinal scenarios, a standardized interface to dynamic test collections is necessary. We extend the \irdatasets framework~\cite{DBLP:conf/sigir/MacAvaneyYFDCG21} with methods tailored to dynamic test collections. We re-implement some of the proposed methods from last year's iteration of the LongEval lab and compare the results in terms of code complexity and retrieval effectiveness.
Our \irdatasets extension provides a valuable benefit for software submissions, as the submitted software can focus on the main ideas of their approach and does not have to do dedicated data wrangling, thereby reducing the maintenance efforts of organizers. Even bugs in the \irdatasets extension can be handled without modifying the code of participants, e.g., by re-installing the extension in the submitted software. We hope that our \irdatasets extension also simplifies software submissions for longitudinal experiments in the future, as submitted software is less complex.

In summary, our core contributions are:
\begin{itemize}
    \item Re-implementation of our original approaches submitted to LongEval 2024 \cite{keller:2025a,alexander:2024,keller:2024}, as PyTerrier transformers~\cite{DBLP:conf/ictir/MacdonaldT20}.
    \item An extension of \irdatasets for longitudinal evaluations.
    \item We added the LongEval Sci and LongEval Web datasets to the extension as the first dynamic test collections.
    \item A preliminary analysis of the re-implemented systems in terms of code complexity and retrieval effectiveness.
\end{itemize}

 The \irdatasets extension\footnote{\url{https://github.com/clef-longeval/ir-datasets-longeval}} and our re-implemented approaches from LongEval~2024\footnote{\url{https://github.com/clef-longeval/longeval-code/tree/main/clef25}} are publicly available on GitHub.
\section{Related Work}
The observation that real-world search engines must operate in an evolving environment motivates longitudinal evaluations. Compared to traditional Cranfield-style evaluations, which use a static set of documents and relevance judgments, longitudinal retrieval evaluations must capture varying versions of relevance judgments and documents, thereby increasing the complexity of the experiments. Streamlining access to dynamic test beds should support researchers in conducting longitudinal evaluations. In this context, we discuss the related work.

\paragraph{Evolving Search Settings.}
While in traditional Cranfield-style evaluations most dynamics are abstracted, in real-world settings each component may change over time and thereby influence the retrieval results~\cite{DBLP:conf/ictir/KellerBS24}. The document corpus or the content of the documents change~\cite{DBLP:conf/wsdm/AdarTDE09,DBLP:conf/infoscale/PassCT06}. Simultaneously, information needs evolve~\cite{DBLP:conf/sigir/Dumais14,DBLP:conf/wsdm/TylerT10,DBLP:conf/infoscale/PassCT06}, ultimately affecting what users perceive as relevant~\cite{DBLP:conf/sigir/AltingovdeOU11}.

\paragraph{Longitudinal Evaluations.} Assessing how well retrieval systems work for users at different points in time requires repeated experiments. This significantly increases the complexity of IR experiments, as every point in time might have different restrictions on what information can be accessed. This evolving setting is investigated by the LongEval CLEF lab~\cite{DBLP:conf/clef/AlkhalifaBBCDEASGGKLLMPPSMZ23,DBLP:conf/clef/AlkhalifaBDEAFGSGILMMMP24} that provides two dynamic test collections with up to 15 snapshots~\cite{DBLP:conf/ecir/CancellieriEFGSGIKKMPPS25,DBLP:conf/sigir/GaluscakovaDSMG23}. Beyond that, not many dynamic evolving test collections are available, so that studies often simulate dynamics or rely on other versioned datasets such as TREC-COVID, with five snapshots~\cite{DBLP:journals/sigir/VoorheesABDHLRS20}.

In many longitudinal experiments, each snapshot requires unique runs. This increases the complexity and computational resources of experiments. For instance, Keller et al. tested five different retrieval approaches on TREC-COVID, LongEval 2023, and a simulated dynamic test collection based on TripClick~\cite{rekabsazTripClickLogFiles2021,DBLP:conf/ictir/KellerBS24}. In total, 45 retrieval runs were created. 
Depending on the simulation strategy, many more snapshots can be created, resulting in even more retrieval runs~\cite{DBLP:conf/clef/ElEbshihyFSGPIGM23}.

\paragraph{Accessing Datasets.}
Thakur et al. propose the BEIR benchmark to investigate the out-of-distribution generalization of IR systems~\cite{DBLP:conf/nips/Thakur0RSG21}. They enable straight forward evaluations over 18 datasets from various domains.  MacAveney et al. introduce \irdatasets, a Python toolkit that provides a standardized interface to datasets typically used in IR research~\cite{DBLP:conf/sigir/MacAvaneyYFDCG21}, which got also integrated into TIRA~\cite{DBLP:conf/ecir/FrobeWKGELHSP23,froebe:2023e} to promote more reproducible shared tasks. The community can contribute to \irdatasets to extend the interface which is essentially what we did for longitudinal evaluations. Currently, the catalogue holds 55 datasets. 
Earlier works that provide interfaces to multi-dataset benchmarks are, for example, MultiReQA with eight datasets and a focus on question answering~\cite{DBLP:journals/corr/abs-2005-02507}.

\section{Interfacing Dynamic Test Collections}
Traditionally, test collections in IR consist of three components: a document corpus, a set of queries, and a set of relevance judgments~\cite{DBLP:journals/ftir/Sanderson10}. In longitudinal evaluations, these three components evolve over time, yielding time as an additional dimension for evaluation. While this evolution is natural to all deployed IR systems, it is challenging to design offline experiments that account for it. Dynamic test collections capture this evolution in the form of versioned sub-collections~\cite{gonzalessaezContinuousEvaluationFramework2023,DBLP:conf/sigir/Soboroff06}. The same test collection is captured repeatedly at different points in time, each with its own document corpus, queries, and qrels. We refer to such sub-collections as a snapshot, as they preserve one point in time of a search setting.
Effective systems in these settings incorporate relevance informations from previous snapshots into their rankings, either by directly relying on previous relevance judgments or by training on previous queries and documents~\cite{DBLP:conf/clef/AlkhalifaBBCDEASGGKLLMPPSMZ23,DBLP:conf/clef/AlkhalifaBDEAFGSGILMMMP24}. We extend the \irdatasets framework~\cite{DBLP:conf/sigir/MacAvaneyYFDCG21} to support dynamic test collections, and provide a unified access to different (subsets of) snapshots and their components.

The \irdatasets toolkit already models the access to datasets using hierarchical dataset IDs. We extend this by directly returning a meta dataset for IDs of higher order. The call for the id \texttt{\small longeval-sci/*} in Listing~\ref{listing-meta-datasets} provides a meta dataset containing all snapshots for the LongEval Sci test collection. 

\begin{listing}
\begin{minted}{python}
from ir_datasets_longeval import load

ir_datasets = load("longeval-sci/*")
for dataset in ir_datasets.get_datasets():
    run_experiment(dataset)
\end{minted}
\caption{Processing all snapshots from the LongEval Sci collection.}
\label{listing-meta-datasets}
\end{listing}

Additional IDs are available for accessing only the snapshots used as test data for a shared task. For example, the ID \texttt{\small longeval-sci/clef2025-test} provides a meta dataset capturing the snapshot \texttt{\small longeval-sci/2024-11} and the snapshot \texttt{\small longeval-sci/2025-01}. This allows for convenient testing of a retrieval approach on all snapshots.

Similar to the \texttt{\small get\_datasets()} method of the meta datasets, we create a \texttt{\small get\_prior\_datasets()} method that returns all prior snapshots of a dataset (Listing~\ref{listing-prior-datasets}). This allows for easy and recursive access to previous qrels and other components, so that no custom logic is needed to ensure approaches only use the allowed parts of the data. The prior snapshots are ordered by their timestamp, so that the most recent snapshot is first. Retrieval approaches can access the most recent snapshots independently of the elapsed time between or the aggregation level of the snapshots.

\begin{listing}
\begin{minted}{python}
for prior_dataset in dataset.get_prior_datasets():
    prior_dataset.get_timestamp()
\end{minted}
\caption{Accessing the prior datasets of a snapshots.}
\label{listing-prior-datasets}
\end{listing}

Listing~\ref{listing-processing-dataset} shows how we extend the dataset object with additional methods to provide metadata relevant to longitudinal evaluations. The \texttt{\small get\_timestamp()} method returns the timestamp of the snapshot. This can be used as a reference point to dynamically calculate the recency of documents. The \texttt{\small get\_snapshot()} method returns the snapshot name to locate a snapshot in the meta dataset. If adequate metadata is available, timestamps for documents can also be retrieved from the document store.

\begin{listing}
\begin{minted}{python}
# At what time does/did a dataset take place?
dataset.get_timestamp()  # 2024-11

# What is the name of the dataset?
dataset.get_snapshot()  # "2024-11"

# When was a document published or updated?
store = dataset.docs_store()
for doc in store.docs_iter():
    doc.publishedDate  # 2016-02-10T00:00:00
    doc.updatedDate  # 2024-02-29T10:01:57
\end{minted}
\caption{Accessing properties of datasets and documents.}
\label{listing-processing-dataset}
\end{listing}

Since the search settings naturally evolve over time, for example when the corpus is updated, frequent changes can be expected. It is desirable to provide the same standardized interface to these newly evolved snapshots without the overhead of registering a new snapshots. Furthermore, retrieval experiments might run on modified versions of the original datasets. To facilitate this, custom, local datasets can be directly loaded from disk. These datasets need to follow the data structure of one of the official dynamic test collections and provide a metadata file with the information about timestamps and prior datasets. Listing~\ref{listing-custom-dataset} shows how a custom dataset can be loaded.

\begin{listing}
\begin{minted}{python}
dataset = load("<PATH-TO-A-DIRECTORY-ON-YOUR-MACHINE>")
\end{minted}
\caption{Loading custom datasets.}
\label{listing-custom-dataset}
\end{listing}

With these extensions, the \irdatasets framework provides a convenient interface to access dynamic test collections. It allows for easy access to previous snapshots and their components, and to use them in retrieval approaches.
\section{Approaches and Re-Implementation}

In the LongEval scenario, we implemented various approaches that utilize prior snapshots of the same test collection to enhance retrieval effectiveness as part of our participation in LongEval 2024. They are based on the heuristic that prior snapshots with their rankings bear some kind of relevance information that can be used for the ranking of the current snapshots. Generally, we derive relevance signals for the ranking at the timestamp $t_n$ from the snapshot at one or more earlier timestamps $t_{n-1}$. Compared to an earlier snapshot, a current ranking at $t_n$ can contain new documents that are added after $t_{n-1}$, documents that were also present at $t_{n-1}$, and documents that were present at $t_{n-1}$ but now have different content. This setting is streamlined with the \irdatasets integration as explained in the previous section. In this work, we focus especially on two approaches that were re-implemented using the \irdatasets extension and PyTerrier transformers~\cite{DBLP:conf/ictir/MacdonaldT20}, namely qrel boosting and relevance feedback.

\subsection{Qrel Boost}
The Qrel Boost approach directly uses the relevance label from a prior snapshot to boost documents that are known to be relevant. We boost documents $d$ for a query $q$ at $t_n$ judged as $rel(q,d,t)$ at a previous snapshot $t_{n-1}$ by:

\begin{equation}
    \text{score}(q,d) \; = \; \text{score}_0 \;\times \;
    \prod_{t = t_1}^{t_k}
    \begin{cases}
        (1 - \lambda)^2, & \text{if } rel(q,d,t) = 0, \\
        \lambda^2, & \text{if } rel(q,d,t) = 1, \\
        \lambda^2  \mu, & \text{if } rel(q,d,t) = 2.
        \end{cases}
\end{equation}

The parameters $\lambda$ and $\mu$ are free and control the general weighting factor and additional weights for highly relevant documents. In the re-implementation, only the dataset needs to be passed to the PyTerrier transformer, examplified in Listing~\ref{listing-qrel-boost}. The prior snapshots are accessed with \texttt{\small get\_prior\_datasets()}. How many prior snapshots should be used can additionally be controlled with the \texttt{\small memory} parameter.

\begin{listing}
\begin{minted}{python}
    pipeline = bm25 >> QrelBoost(dataset, memory=1)
\end{minted}
\caption{QREL Boost as PyTerrier Transformer applied to BM25.}
\label{listing-qrel-boost}
\end{listing}

\subsection{Relevance Feedback}
Our relevance feedback approach expands the queries with terms from previously relevant documents. Each query is expanded with the top k terms with the highest tf-idf scores. In the original implementation, the tf-idf scores were calculated using an additional database with all document texts. With the proposed \irdatasets extension, the document texts can be directly accessed through the \irdatasets document store. Still, the re-implementation relies directly on the tf-idf scores from the index of the previous snapshot since this is a theoretically better indicator for relevant terms. The directory of all prior indices is passed to the PyTerrier transformer (see Listing~\ref{listing-rf}) and the correct index is identified by the \texttt{\small get\_snapshot()} method.

\begin{listing}
\begin{minted}{python}
    indices = "path/to/indices"
    pipeline = RF(dataset, indices, memory=1) >> bm25
\end{minted}
\caption{Relevance Feedback as PyTerrier Transformer passed as query reformulation into BM25.}
\label{listing-rf}
\end{listing}

\section{Experiments and Evaluation}
As an experimental evaluation of our contributions, we reimplement our submissions to LongEval~2024 with our new \irdatasets extension and measure the code complexity and the reproduction of retrieval results.

\subsection{Code Complexity}
To analyze how the \irdatasets extension affects implementation and experimentation efforts of longitudinal retrieval experiments, we analyze the code complexity of the implementation of the retrieval approaches. Therefore, the initial version of the approach from~\cite{keller:2025a} and its new re-implementation are compared using the code complexity measures from the Lizard toolkit.\footnote{\url{https://github.com/terryyin/lizard}} %
Lizard provides metrics such as the total number of lines of code (NLOC), average NLOC per function (Avg.NLOC), the average cyclomatic complexity (AvgCCN), average number of tokens per function, and the number of functions (Fun Cnt). The Average NLOC is the total number of lines of code divided by the number of functions, and the Average CCN is the total cyclomatic complexity divided by the number of functions. The AvgCCN measures the average number of independent paths through the functions in the code. A lower value indicates simpler, more maintainable functions with fewer decision points~\cite{DBLP:journals/tse/McCabe76}.

A direct comparison of the code bases is difficult as the initial version is scattered across at least three files, with two of them shared. Additionally, a database was used to access document information that is no longer needed. Still, the results generally indicate a decreased complexity for the re-implementations. Table~\ref{tab:code_metrics_summary} provides an overview of the complexity measures.

For both approaches, the total NLOC and the Avg. NLOC is decreased. The AvgCCN is generally low, which indicates maintainable code in both cases (which is reasonable as the ideas are conceptually easy). However, it slightly increased for the qrel\_boost approach, but decreased for relevance\_feedback. Therefore, the new version of the relevance\_feedback approach gained two more functions while the qrel\_boost approach lost five functions.

These results indicate a generally decreased complexity. This is further supported by the fact that they now better adhere to PyTerrier and \irdatasets, two widely adopted frameworks in the community.
\begin{table}[t]
\centering
\caption{Code complexity metrics for the re-implemented approaches. The first row of each group shows the initial version of the approach from~\cite{keller:2025a} and the second row shows the new re-implementation. The best values per group are highlighted in \textbf{bold}.}
    \setlength{\tabcolsep}{3pt}

    \begin{tabular}{r | c  c  c  c  c }
    \toprule
    {} & Total NLOC & Avg. NLOC & AvgCCN & Avg. token & Fun Cnt \\
    \midrule
    qrel\_boost & 250 & 17.8 & \textbf{2.6} & 142.2 & 12  \\
    qrel\_boost new & \textbf{99} & \textbf{12.0} & 2.9 & \textbf{98.9} & 7 \\
    \hline
    relevance\_feedback & 231 & 23.8 & 3.0 & 170.6 & 8 \\ 
    relevance\_feedback new & \textbf{197} & \textbf{17.3} & \textbf{2.5} & \textbf{124.8} & 10 \\
    \bottomrule
    \end{tabular}
    \label{tab:code_metrics_summary}
\end{table}

\subsection{Replication of Retrieval Results}
Finally, we want to ensure that the re-implementations of the approaches still achieve similar retrieval results to the original approaches. Currently, only the most recent LongEval datasets are added to the extension because they also contain all snapshots from 2023 and 2024. However, this year's version processed the data differently and also contained many more queries. For example, in the 2024 version of the dataset, new document IDs were assigned for each new version of a website. This is unified in the current dataset, which better supports temporal approaches that rely on prior snapshots. Given these changes, a direct comparison is difficult, and the focus needs to be on the general conclusions instead of absolute results. 

We measured how well the re-implementations replicate the results of the original approaches according to the Delta Relative Improvement ($\Delta$RI), Effect Ratio (ER) and the p-values of unpaired t-tests as proposed by Breuer et al.~\cite{DBLP:conf/sigir/Breuer0FMSSS20}. To compare the runs of different topic sets across implementations, with $\Delta$RI and ER are the runs related to a reference system, BM25 in our case. Only the deltas are then compared across the original and the re-implemented approaches. 
The $\Delta$RI describes the relative change in effectiveness of the re-implementation compared to the original approach. The further $\Delta$RI diverges from zero, the weaker the replication.
The ER measures how well the per-topic difference from the reference system is recovered in the re-implementation. The closer the ER is to one, the better the re-implementation recovers the differences from the reference system.
The p-value of the unpaired t-test indicates whether the difference between the original and re-implementation is statistically significant. Since we aim to achieve similar results from the re-implementation, no significant differences with p-values over 0.05 are desirable. The experiments were carried out with \texttt{repro\_eval}~\cite{DBLP:conf/ecir/BreuerFMS21}.

\begin{table}
\caption{This table reports the replicability results based on $\Delta$RI, ER, and $p$-values. Additionally, the average retrieval performance is reported for the re-implementation and also the original version in \textcolor{teal!90}{teal}. All results are based on nDCG@10.}
\label{tab:replication}
\begin{center}
\setlength{\tabcolsep}{8pt}
\begin{tabular}{lccccccccc}
\toprule
snapshot & system & ER & $\Delta$RI & $p$-value & nDCG@10 \\
\midrule
2022-07 & qrel\_boost & 2.307 & -0.407 & 3.63e-13 & \textcolor{teal!90}{0.245}/0.343\\
2022-07 & relevance\_feedback & 3.230 & -0.535 & 8.91e-20 & \textcolor{teal!90}{0.229}/0.351 \\
\midrule
2022-09 & qrel\_boost & 1.943 & -0.458 & 2.25e-05 & \textcolor{teal!90}{0.259}/0.317 \\
2022-09 & relevance\_feedback & 2.224 & -0.461 & 5.90e-06 & \textcolor{teal!90}{0.243}/0.303 \\
\midrule
2023-01 & qrel\_boost & 4.925 & -0.522 & 5.24e-36 & \textcolor{teal!90}{0.234}/0.445 \\
2023-01 & relevance\_feedback & 3.685 & -0.312 & 1.30e-21 & \textcolor{teal!90}{0.231}/0.386 \\
\midrule
2023-06 & qrel\_boost & 1.569 & 0.003 & 3.49e-13 & \textcolor{teal!90}{0.268}/0.421 \\
2023-06 & relevance\_feedback & 1.723 & -0.028 & 1.65e-11 & \textcolor{teal!90}{0.220}/0.355  \\
\midrule
2023-08 & qrel\_boost & 1.896 & -0.071 & 1.75e-42 & \textcolor{teal!90}{0.199}/0.338 \\
2023-08 & relevance\_feedback & 2.018 & -0.050 & 7.46e-34 & \textcolor{teal!90}{0.171}/0.287 \\
\bottomrule
\end{tabular}
\end{center}
\end{table}

The replication results based on nDCG@10 are reported in Table~\ref{tab:replication}. They indicate that the re-implementation most often achieves different results. This is most likely due to the differences in the collections. Given the low p-values, the results are significantly different. The ER always exceeds the ideal value of 1, meaning that the re-implementations are not replicating the same effect but achieve higher differences compared to the reference systems. This is further supported by the almost always negative $\Delta$RI. Additionally, the scores indicate that the relative improvement of the overall effect is relatively well replicated, especially on the later snapshots.

\section{Conclusion}
We describe how longitudinal retrieval experiments can be simplified with an extension for \irdatasets that improves the support for dynamic test collections. We demonstrated its utility by re-implementing previous approaches as PyTerrier transformers that make use of the added functionality. These changes simplify software submissions for the LongEval lab, piloting this year with the SciRetrieval task.
We examined how the proposed changes affect longitudinal experiments in terms of code complexity and how well the re-implementations replicate the original retrieval effectiveness. The results indicate that the overall complexity decreased. Although the initial results could be only weakly replicated, the effectiveness of the approaches improved. This is most likely due to the updated test beds and the improvements made to the approaches.

Future works regard improvements and extensions of the proposed methods.
The \irdatasets extension for dynamic test collections is a first step towards a standardized interface to longitudinal evaluations. While not many dynamic test collections are available yet, new datasets should be added as they become available. Furthermore, related datasets with versioned collections could be added, such as Soboroff's evolving version of the GOV2 dataset~\cite{DBLP:conf/sigir/Soboroff06}.
A limitation of the extension is that each snapshot requires a complete document corpus, even if it is only partially changed. This could be addressed in the future by implementing approaches that only store the deltas between snapshots, similar to~\cite{DBLP:conf/sigir-ap/StaudingerPR24}.
Finally, it needs to be checked what features can be directly integrated into the \irdatasets framework and what should be kept as a separate extension. 

With the proposed extension, we hope to streamline longitudinal experiments and help researchers to cope with the additional complexity that longitudinal experiments introduce.

\begin{credits}
\subsubsection{\ackname}
We gratefully acknowledge the support of the German Research Foundation (DFG) through project grant No. 407518790.
\subsubsection{\discintname}
The authors Jüri Keller, Maik Fröbe, and Philipp Schaer joined the LongEval organization team in 2025. All other authors have no competing interests to declare that are relevant to the content of this article.
\end{credits}

\bibliographystyle{splncs04}
\raggedright
\bibliography{clef25-longeval-invited-lit}

\end{document}